# Fouling Mitigation in Tubular Membranes by 3D-printed Twisted Tape Turbulence Promoters

Sarah Armbruster[a], Oskar Cheong[a], Jonas Lölsberg[a,b], Svetlana Popovic[c], Süleyman Yüce[a], Matthias Wessling[a,b,*]

[a]*RWTH Aachen University, Aachener Verfahrenstechnik - Chemical Process Engineering, Forckenbeckstr. 51, 52074 Aachen, Germany*
[b]*DWI - Leibniz Institute for Interactive Materials, Forckenbeckstr. 50, 52074 Aachen, Germany*
[c]*Department of Chemical Engineering, Faculty of Technology, University of Novi Sad, Bulevar cara Lazara 1, 21000 Novi Sad, Serbia*

**Abstract**

Despite intensive research, fouling remains a severe problem in membrane filtration. It is often controlled by applying turbulent flow which requires a higher energy consumption. So-called turbulence promoters or static mixers can be inserted into the flow channel of tubular membranes. They deflect the fluid, induce vortices, enhance particle back-transport and increase the shear rate at the membrane surface, thus mitigating fouling. However, little is known how the geometry of such turbulence promotors affects the reduction of fouling. We investigate how different 3D-printed mixer geometries affect fouling and improve the flux during filtration with humic acid. Most mixer geometries used in the present study are based on a twisted tape; a Kenics static mixer is investigated as well. Static mixers with changing diameter prove to be less effective than twisted tape mixers with constant diameter which lead to an increase in permeate flux of around 130 %. The highest flux improvement of 140 % can be reached by applying a Kenics mixer. Regardless of their geometry, all investigated static mixer cause higher permeate fluxes at same specific energy consumption. Again, the Kenics mixer proves to be the most efficient static mixer. The presented mixer geometries can be fabricated with undercut injection molding techniques and represent a simple and viable option to make tubular membrane based filtration processes more efficient.

[*]*Corresponding Author: M. Wessling*
   *Email address:* `manuscripts.cvt@avt.rwth-aachen.de` (Matthias Wessling)

---

## 1. Introduction

Membrane filtration has long been used in a broad variety of industries ranging from the food and beverages industry to water treatment for all kinds of (waste)water to purifying highly valuable product streams in biotechnology. In contrast to other separation technologies, membrane filtration does not require large amounts of energy. This is one of the major benefits of membrane processes and makes them a sustainable, energy- and cost-efficient alternative to traditional thermal separation processes. Despite some progress, concentration polarization and fouling remain challenges in membrane filtration. The term fouling comprises several phenomena such as adsorption, pore constriction, pore blocking, cake formation, and the formation of biofilms. Regardless of the actual mechanism, all fouling phenomena reduce the permeate flux significantly when operating at constant transmembrane pressure (TMP) and increase the TMP when using constant flux-mode, respectively. Once fouling has occurred and the membrane's filtration capacity has declined, the membrane has to be cleaned or even replaced. However, cleaning is costly as not only chemicals might be needed, but also the filtration is interrupted. Hence, the main focus has to be put on mitigating fouling to avoid cleaning interruptions of filtration processes that would weaken the economic viability of membrane processes.

Numerous methods have been tested to reduce fouling. One promising approach to mitigate fouling is to alter the hydrodynamics in the module. Hydrodynamic fouling countermeasures usually aim at increasing the shear rate at the membrane surface and at increasing the particle back-transport into the bulk [1, 2]. An easily applicable fouling countermeasure is to use turbulent flow conditions. Turbulence reduces the laminar boundary layer thickness, increases the shear rate at the membrane surface and causes mixing perpendicular to the membrane surface. These phenomena help to reduce fouling as foulants cannot deposit as easily as in laminar flow. Turbulent flow conditions are mostly achieved by increasing the cross-flow velocity. However, higher cross-flow velocities result in a higher energy consumption. Another method to increase the wall shear rate, enhance particle back-transport and reduce the boundary layer thickness is the application of static inserts in the membrane flow channel. In literature, they are referred to as static mixers or turbulence promoters [1–4]. As their names imply, they deflect the fluid, induce secondary flows such as vortices and enhance mixing [5–7]. An overview about the application of various kinds of static mixers and of spacers for flat sheet membranes is given by Popovic *et al.* [2].

Static mixers have long been used in heat exchangers, as a disruption of the boundary layer, higher shear rates, mixing, and secondary flows increase heat transfer [8–10]. A comparison of different types of static mixers regarding their influence on heat transfer was carried out by Yüce [11], for instance. Sulzer SMX, SMX-L, Kenics and twisted tape static mixers were applied to increase the heat transfer in several visco-elastic fluids. It was shown that an improvement in heat transfer is always accompanied by a higher pressure loss. Regarding the achieved heat transfer enhancement, Kenics mixers were more effective than twisted tape turbulence promoters [11]. On the other hand, the Kenics mixers caused a higher pressure loss than twisted tape turbulence promoters. This tendency is also reflected by the performance number, which relates the exchanged heat flow per unit of the driving temperature difference to the dissipated pump power. A comparison of various static mixers based on this performance number showed that those mixers which deflect the flow only slightly (e.g. twisted tape) are energetically better than static mixers with more hydrodynamic influence as the Kenics mixer [11]. This finding is more significant for highly viscous polymer solutions [11]. High viscosity and high viscoelastic fluid properties reduce the Reynolds number and lead to a smoother flow pattern. Therefore, heat transfer is significantly decreased for highly viscous fluids. This is why static mixers are particularly for viscous fluids a very effective means for improving heat transfer [11]. Based on these findings in the field of heat transfer, similar effects of static mixers on mass transfer are expected.

For this reason, static mixers have been investigated in membrane filtration [1, 3, 4, 12, 13]. Several types of inserts have been used; ranging from simple inserts such as rods [1, 14, 15] over twisted tapes [4] to more complex structures such as Kenics mixers [12, 16–18]. All of them have in common that they obstruct the flow and thereby lead to higher cross-flow velocities in the tubular flow channel of the membrane [1]. In addition, higher cross-flow velocities increase the wall shear rate at the membrane surface [12]. To further enlarge the benefit of inserts in the flow channel, their geometry is designed to induce secondary flows which enhance mixing and particle back transport into the bulk fluid [4, 19]. For instance, wires are wound around a rod to establish a helical flow path forming vortices and flow instabilities [7, 20]. Several pitch lengths or turns of the wire per rod length have been tested [7, 20–22]. With this simple method, Ahmad *et al.* [7] found flux improvements of up to 100 % in the microfiltration of $TiO_2$ particles. A variation of wired rods are helical screw inserts as, for instance, Liu *et al.* [23] used in the microfiltration of carbonate suspensions. Another type of screw-threaded inserts were applied in the plasmapheresis of bovine blood [24]. Mavrov *et al.* [25] changed the geometry of a smooth rod by varying its cross-section and using cone-shaped repeating units. In comparison to a smooth rod, the cone-shaped insert led to higher fluxes at the same specific energy consumption. This was explained by the change in flow path induced by the varying cross-section of the cone-shaped insert [25].

A helical flow path can also be reached by inserting twisted tapes in the flow channel of tubular membranes. Popovic *et al.* [4] investigated the influence of twisted tape turbulence promoters with different aspect ratios on the microfiltration of milk. The tightest twisted tape with the lowest aspect ratio was found to have the highest flux improvements of up to 600 % at the same cross-flow rate. The higher efficiency of the lower aspect ratios was attributed to the longer flow path and hence higher velocity of the fluid due to the increased number of twists per length [4].

Another variation of static mixers are blade-type inserts with open space in the middle of the mixer. The open space in the tube center should decrease the pressure loss of the mixer and thus require less energy compared to mixers which also obstruct the center of the flow channel. Popovic *et al.* [19] investigated two variations of blade-type mixers with different aspect ratio in the microfiltration of skim milk. To reach a flux of 80 $LMH$, the system without mixers required 6 $kWh/m^3$, whereas less than 4.4 $kWh/m^3$ were necessary in the system with blade-type mixers [19]. However, the specific energy consumption was even lower when twisted tape mixers were applied instead of blade-type mixers [19]. The lower energy-efficiency of the blade-type mixers in comparison to the twisted tape mixers was attributed to the higher pressure loss the blade-type mixers impose on the filtration system [19].

The mixing abilities of the Kenics mixer are well-known [9, 12, 17, 18, 26, 27]. A Kenics mixer consists of alternating left- and right-twisted tapes which are displaced by 90° to each other. Krstic *et al.* [16] applied Kenics mixers in cross-flow filtration of skim milk and found flux improvements of over 700 %. Although the pressure drop along the membrane rose by about 300 %, the specific energy consumption was reduced by around 29 % when a Kenics mixer was applied [16]. Similar results with flux improvements of up to 600 % and a 40 % reduction in specific energy consumption were found in Kenics-assisted ultrafiltration of an oil-in-water emulsion [17]. A Kenics static mixer has also been applied in the microfiltration of yeast suspensions and improved the permeate flux by 90 − 260 % [18]. By means of response surface methodology, the authors found that transmembrane pressure and feed flow rate had the most significant influence on the permeate flux when a Kenics mixer was used in the microfiltration of yeast [18]. Gaspar *et al.* [28] compared a single tubular ceramic membrane fitted with a Kenics mixer to a module equipped with ceramic capillary membranes. At lower cross-flow velocities, the single membrane with Kenics mixer required less specific energy for the filtration of an oil-in-water emulsion than the capillary module. These studies show that a Kenics mixer is well-suited to improve the filtration in various systems. Nevertheless, its actual effectiveness depends on the specific foulant involved.

This study aims at comparing the effectiveness of various static mixer geometries on membrane fouling. We compare the well-known Kenics mixer to known and new mixer geometries which are based on a twisted tape geometry. The new geometries include twisted tapes with corrugated edges, twisted tapes with constant diameter and twisted tapes whose diameter varies with the length. In fact, we demonstrate that 3D-printing can be used to fabricate complex static mixers with unprecedented geometries. Thus, the influence of the edge architecture and of the mixer diameter can be assessed. Furthermore, the pitch length or aspect ratio of the mixers is varied in this study. A shorter pitch length leads to a longer pathway for the fluid and hence deflects it more strongly [4], however, it increases the pressure loss as well. Evaluating the specific energy required per $m^3$ permeate allows to assess the benefits of static mixers in filtration.

## 2. Materials and Methods

Fouling experiments were performed with an aqueous solution of 20 $mg/L$ humic acid. The feed solution was prepared by dissolving humic acid sodium salt by Carl Roth GmbH + Co.KG in de-ionized water. The $ZrO_2$ ultrafiltration membrane used was provided by atech innovations GmbH, Germany, and has a nominal cut-off of 100 $kDa$. The same single-channel, tubular membrane was used for all experiments and cleaned in between single experiments with 0.01 $M$ or 0.1 $M$ NaOH, depending on the extent of fouling. Its flow channel has an inner diameter of 6 $mm$ and a length of 250 $mm$. The effective filtration area was 0.00413 $m^2$.

*Turbulence Promoters*

Nine different turbulence promoters were investigated in this study (see Fig. 1). They were manufactured in-house by polyjet 3D printing (Stratasys, Objet Eden 260V) and consist of a photosensitive acrylate-based polymer (Stratasys, RGD810). A second polymer (Stratasys, SUP705) is necessary to support the static mixers during the printing process. After printing, the supporting structure was removed in a 1 $M$ NaOH solution.

Eight of the turbulence promoters are based on the twisted tape geometry, whereas the ninth is a Kenics static mixer. All of them were 230 $mm$ long, had a maximum diameter of 5.5 $mm$, a minimum diameter of 3 $mm$ and a thickness of 1 $mm$. Except the Kenics mixer, each mixer geometry was fabricated in two versions with differing pitch length (13.75 $mm$ and 8.25 $mm$). The pitch length is defined as promoter length divided by the number of twists. Hence, it corresponds to the length of one twisted element of the promoter.

The investigated geometries of the turbulence promoters are shown in Fig. 1. The uppermost static mixer is a plain twisted tape. In this study, it is referred to as twisted tape or "Helix". Directly beneath, the statix mixer with "dotted" structure is shown. Its geometry is directly based on the twisted tape, but has a structured edge to increase the formation of vortices. The so-called sinusoidal turbulence promoter is based on a tape with a straight width and a sinusoidal length instead of two straight sides. Hence, its diameter continuously changes from 5.5 $mm$ to 3.5 $mm$ and back to 5.5 $mm$. Twisting this tape results in the sinusoidal mixers displayed in position 5 and 6 in Fig. 1. The next two mixers shown in Fig. 1 are the "step helix" promoters. Their diameters are also altering along their axis. For both pitch lengths, the diameter decreases over the length of 3 twisted elements from 5.5 $mm$ to 3.5 $mm$. Then, the diameter increases sharply back to 5.5 $mm$, leading to a pronounced edge in the promoter's geometry. The Kenics static mixer consists of right-turning and left-turning twisted tape elements which are attached to each other in altering order at an angle of 90° (see last mixer in Fig. 1). Each twisted element has a pitch length of 13.75 $mm$.

The pure water permeability of the ceramic membrane was measured in dead-end in the range of 0.5 $bar$ to 1.5 $bar$. It was re-measured in between two fouling experiments to ensure sufficient cleaning. For the fouling experiments, the set-up shown in Fig. 2 was used. The system consists of a gear pump, pressure and flow sensors, the membrane module, a feed tank and a weighing scale to measure the permeate flow rate. During the experiments, the feed flow rate was kept constant at 30 $L/h$. This correlates to a Reynolds number of 1765 in the empty flow channel of the tubular membrane. The pressures of feed, retentate and permeate were continuously measured, as well as the retentate and permeate flow rates.

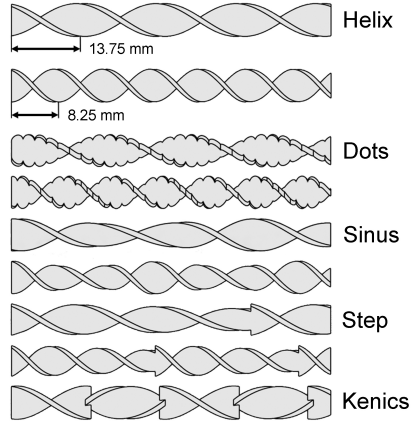

Figure 1: Investigated static mixers. Except the Kenics mixer, each static mixer geometry was fabricated with two different pitch lengths of 13.75 $mm$ and 8.25 $mm$. The pitch length corresponds to the length of one twisted element.

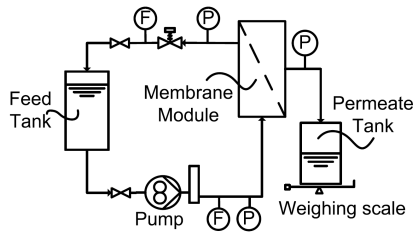

Figure 2: Flow sheet of the set-up used for the fouling experiments.

A pressure-step method was applied to assess fouling as described by Espinasse *et al.* [29]. The transmembrane pressure (TMP) is alternatingly increased and decreased. Each TMP is kept until the measured flux reaches a constant value which leads to varying durations for the single TMP steps (see

Fig. 3). In Fig. 3, two exemplary fouling experiments are shown. Left, the system without static mixer is displayed, whereas right, the course of experiment for a system with static mixer is shown. As the mixer reduces fouling, it takes less time to reach the steady-state. Therefore, individual pressure steps are applied shorter in the system with mixer than in the system without mixer.

As the TMP is alternatingly increased and decreased, each TMP is applied twice. Hence, two flux values are obtained per TMP value. These two flux values are the same if there is no fouling or if the fouling is completely reversible by pressure relaxation [29]. In case of (irreversible) fouling, the flux will be lower when a certain TMP is reached the second time. The reason for this is the build-up of a fouling layer during the filtration at higher pressures which are applied in between two steps of the same TMP value (see Fig. 3) [29]. Due to this fouling layer the flux measurements are more reproducible for the second time a pressure step is applied. Therefore, those values are displayed in the flux-TMP graphs.

All experiments were carried out in triplicates. The results shown are averaged values and the 95 % confidence interval of the mean is displayed as error bars.

## 3. Results and Discussion

### 3.1. Influence of Pitch Length and Corrugated Edges

Fig. 4 shows the results of fouling experiments with (a) helically shaped static mixers and (b) helical static mixers with a dotted structure on the edges. Both types of turbulence promoters only differ in the structure (or non-structure) on the edges. The helically shaped mixer is also called twisted tape turbulence promoter. As Fig. 4 clearly depicts, the application of static mixers drastically increases the permeate flux at same TMP. In comparison to the system without mixer where a flux of 105 $LMH$ is reached at a TMP of 1.3 $bar$, the systems with a twisted tape static mixer achieve fluxes as high as 240 $LMH$ at the same TMP.

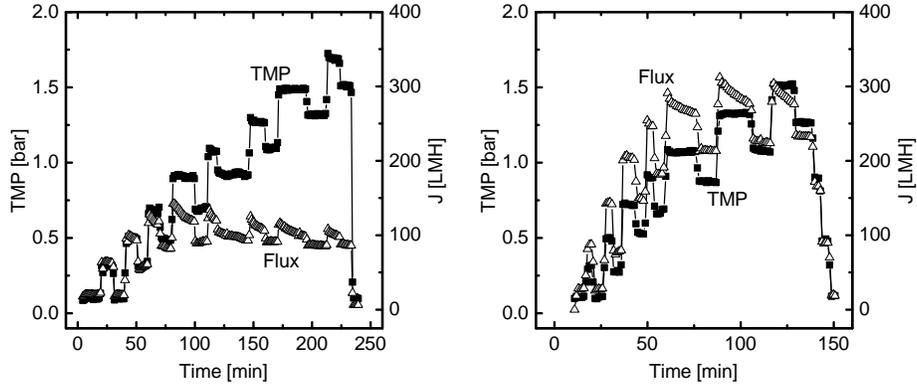

(a) Fouling experiment for a system without static mixer.

(b) Representative fouling experiment for a system with static mixer (helix with a pitch length of 13.75 $mm$).

Figure 3: Pressure-step method: The TMP (filled squares) is alternatingly increased and decreased and the flux (open triangles) measured. Two representative experiments for (a) a system without static mixer and (b) a system with static mixer are shown. Note the different time scales.

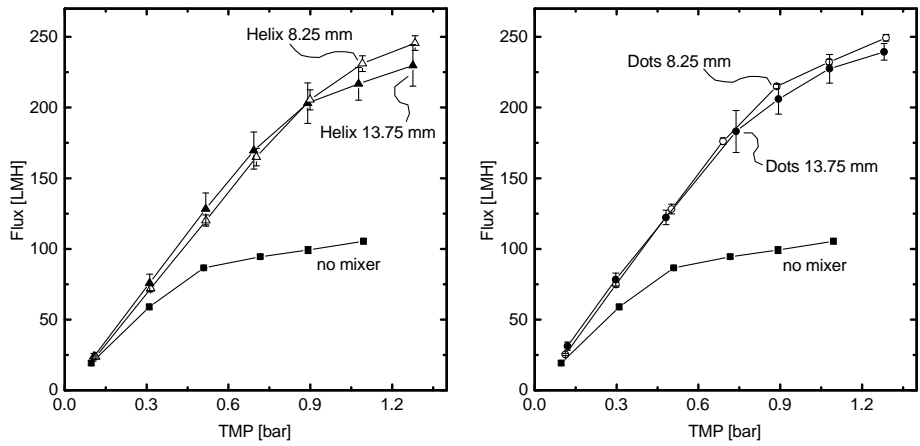

(a) Helically shaped static mixers with pitch lengths of 13.75 $mm$ and 8.25 $mm$, respectively.

(b) Helically shaped static mixers with corrugated edges and pitch lengths of 13.75 $mm$ and 8.25 $mm$, respectively.

Figure 4: Flux vs TMP for a system without mixer and systems with twisted tape static mixers.

Within each subfigure of Fig. 4, the influence of the so-called pitch length is visible. The pitch length basically corresponds to the inverse of the number of twists. Hence, the static mixer with a pitch length of 13.75 $mm$ consists of less twists than the mixer with a pitch length of 8.25 $mm$. As Fig. 4 shows, the pitch length influences the fouling behavior at high fluxes only. The more twisted a static mixer is, the higher the flux at a certain TMP. This is expected as more twists lead to more vortices, a higher pressure loss and hence a better mixing of the fluid [4]. However, the difference between the helically shaped promoter with the lower number of twists (pitch length 13.75 $mm$) and the mixer with the same geometry, but higher number of twists (pitch length 8.25 $mm$) is relatively small. The static mixer with a pitch length of 13.75 $mm$ displays a flux of 230 $LMH$ at a TMP of 1.3 $bar$, whereas the mixer with the lower pitch length shows a flux of 246 $LMH$ at the same TMP. Comparing the two pitch lengths of the corrugated promoter with each other, the difference here accounts for around 10 $LMH$ only at the highest TMP of 1.3 $bar$ and may not even be statistically significant.

The corrugated edges seem to have only minor influences on the fouling behavior of the membrane system when compared to the "plain" twisted tape mixer. At a TMP of 1.3 $bar$, the difference between the two mixers with a pitch length of 13.75 $mm$ accounts for 9 $LMH$, whereas the difference between the two mixers with the lower pitch length is 3 $LMH$ and thus negligible. The corrugated edge was introduced to further increase the active membrane area and to facilitate the formation of additional vortices. As the fouling mitigation of the structured and the "plain" twisted tape is quite similar, the "dots" seem to lead to less additional vortices than expected. However, under different conditions such as a lower Reynolds number, a different flow pattern could evolve and the structured twisted tape might prove to be superior in a slow, laminar flow.

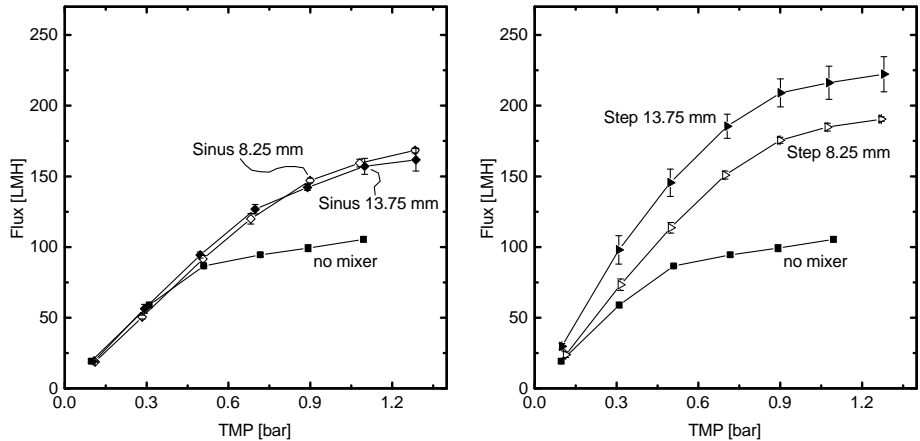

(a) Sinusoidal static mixers with pitch lengths of 13.75 $mm$ and 8.25 $mm$, respectively.

(b) Static mixers with a step and pitch lengths of 13.75 $mm$ and 8.25 $mm$, respectively.

Figure 5: Flux vs TMP for a system without mixer and systems with static mixers whose diameter decreases and increases continuously (sinusoidal shape) or decrease continuously to increase in a discrete step.

*3.2. Static Mixers With Decreasing Diameter*

In addition to twisted tapes with a constant diameter, two mixer geometries where the mixer diameter changes with length were investigated. These are the so-called "step helix" mixer and the sinusoidal static mixer (see Fig. 1). For both mixers, the flow direction was from left to right in Fig. 1, hence in direction of decreasing diameter. Compared to the system without static mixer, the application of the sinusoidal and the step helix mixer improve the filtration significantly (Fig. 5). At a TMP of 1.1 *bar*, the flux is about 60 % and more than 120 % higher for the systems with the sinusoidal and the step helix promoter, respectively, when compared to the system without mixer. The application of turbulence promoters with a constant diameter, though, leads to even higher flux improvements (cf. Fig. 4).

As for the twisted tape mixers with constant diameter, also two different pitch lengths of the mixers with changing diameter were investigated. Comparing the two pitch lengths for the sinusoidal promoter to each other, applying a lower pitch length and thus a more twisted mixer leads to slightly higher fluxes. This finding is the same as for the turbulence promoters with constant diameter. It can be explained by the stronger deflection and redirection of the fluid due to the higher amount of twists [4].

However, the step helix mixer shows a contradictory behavior. Here, the static mixer with the higher pitch length of 13.75 $mm$, hence the less twisted one, leads to higher permeate fluxes than the application of the mixer with lower pitch length (8.25 $mm$). This behavior is unexpected. It might be explained by the steepness in the change of diameter. As for both pitch lengths the diameter of the mixer changes within 3 elements from 5.5 $mm$ to 3.5 $mm$, the decrease in diameter is much steeper for the mixer with more twists (pitch length 8.25 $mm$). Hence, the static mixer with the lower pitch length obstructs the flow channel inside the tubular membrane less and short-cut streams might evolve which are not deflected at all. These phenomena could explain why the more twisted step helix mixer (pitch length 8.25 $mm$) with the steeper decrease in diameter enhances the filtration not as much as its less twisted counterpart (pitch length 13.75 $mm$). CFD simulations or flow visualization methods could be applied to confirm these assumptions.

*3.3. Influence of the Asymmetry of Static Mixers on Fouling Mitigation*

As the step helix is an axially asymmetrical static mixer, it can be applied in two different flow directions. Variation 1 (*var1*) represents the flow against the step of the turbulence promoter, hence the diameter of the mixer decreases continuously and increases in a sharp "step" in flow direction (cf. the second- and third-last mixers in Fig. 1). In the second variation (*var2*), the mixer is turned by 180°, so that the feed flows "in direction" of the step. Thus, the diameter of the static mixer increases continuously and decreases sharply in flow direction. Both flow variations were investigated.

Fig. 6 clearly shows a difference in fouling mitigation depending on the direction in which the asymmetric step helix mixer is applied. Variation 1, where the feed flows "against" the step, led to a significantly higher flux than if the promoter was turned around (*var2*). At a TMP of 1.3 *bar*, variation 1 reached a flux of 225 $LMH$, whereas variation 2 led to a flux of 180 $LMH$ at the same pressure. Qualitatively, the same results were obtained for the step helix mixer with the lower pitch length of 8.25 $mm$ (data not shown). In variation 1, the feed flow is directed towards the obstacle and is deflected by a sharp increase in the mixer's diameter. Thus, the flow is more strongly disturbed and disrupted which then leads to more vortices, a higher shear-rate and better mixing. Therefore fouling is decreased and the permeate flow rate enhanced. All these phenomena might be less pronounced for variation 2, where the feed flow was not directed against the step, but the mixer's diameter increased continuously and decreased sharply. This might explain the difference in flux enhancement for the two variations how to apply the step helix mixer. Nevertheless, both variations strongly mitigate the fouling when compared to the system without static mixer (Fig. 6).

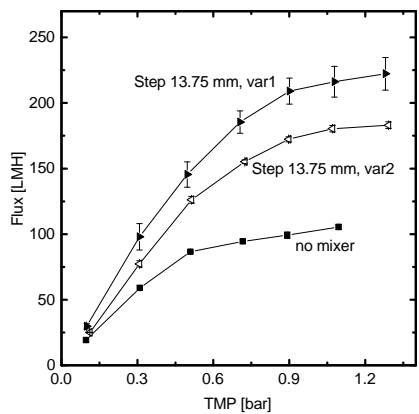

Figure 6: Flux vs TMP for a system without static mixer and systems with "step helix" static mixer. *Var1* denotes a flow direction "against" the step, i.e. in direction of the decreasing diameter. The flow direction was reversed for *var2*, where the fluid flows in direction of the increasing diameter, but is not obstructed by the step.

*3.4. Comparison of Kenics Mixer and Twisted Tape Static Mixer*

In comparison to the system without static mixer, the system with the Kenics mixer showed much higher fluxes, irrespective of the applied TMP (see Fig. 7). For the system with Kenics mixer, the flux does not level off within the studied

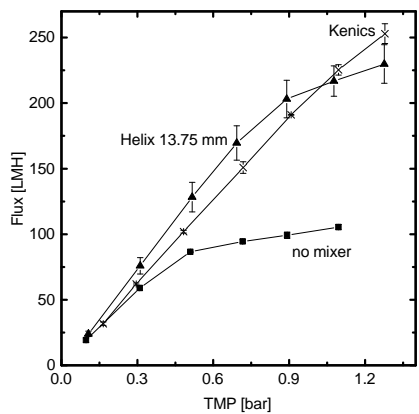

Figure 7: Flux vs TMP for a system without static mixer and systems with a helically shaped static mixer and a Kenics mixer, respectively. For the Kenics mixer, the flux curve merely deviates from a linear line.

TMP range. At the highest applied TMP of 1.3 $bar$, the flux improvement of the Kenics mixer was 140 %. This was the highest flux improvement of all static mixer geometries which were investigated in this study. Even higher flux improvements have been reported: Krstic *et al.* [17] found up to six times higher permeate fluxes in the filtration of oil-in-water-emulsions when applying a Kenics static mixer. However, it has been shown in literature that the actual extent of fouling mitigation and increase in permeate flux strongly depends on the investigated foulant [16–18, 26].

Unlike the fouling curves of all other investigated systems with static mixers, the fouling curve of the Kenics stayed nearly linear in the whole range of TMP analyzed (0.1 $bar$ to 1.3 $bar$; see Figs. 4, 5 and 7). This indicates that fouling starts occuring at a TMP higher than 1.3 $bar$ when a Kenics mixer is applied. Hence, the application of a Kenics mixer would further increase the range of TMP in which the system can be operated without suffering from severe fouling.

*3.5. Specific Energy Consumption and Pressure Drop*

Fig. 8 displays the pressure loss of selected static mixers at different flow rates. These pressure drops were measured with pure water in pipe flow, i.e. no filtration took place. The shown pressure drop corresponds to the difference of inlet and outlet pressure.

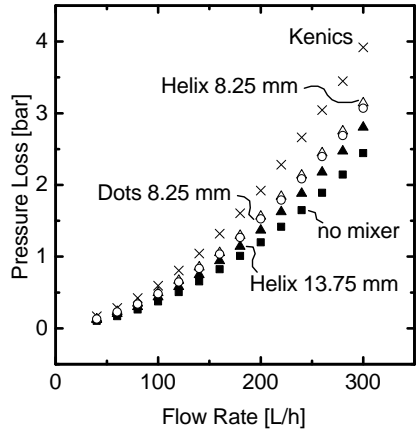

Figure 8: Pressure loss of the system without mixer and selected systems with static mixer at different volumetric flow rates.

Fig. 8 clearly shows that the pressure loss of a mixer depends on its geometry. The mixers with the lower pitch length display a higher pressure drop than the mixers with a higher pitch length. This is attributed to the stronger deflection of the fluid due to the more twisted mixers (shorter element length means more twists per length). Even the corrugated edge has a measurable influence on the pressure drop of the mixer as can be concluded from the difference in pressure drop of the helix mixer and the "dots" mixer (each with 8.25 $mm$ pitch length). On average, the pressure loss imposed by the "dots" mixer with a pitch length of 8.25 $mm$ is 27 % higher than the pressure loss of the pipe flow without static mixer. In contrast, the pressure loss of the helix mixer with a pitch length of 8.25 $mm$ is on average 30 % higher than the pressure loss of the pipe.

Of all investigated mixer geometries, the Kenics mixer deflects the fluid the strongest. Hence, it was expected that the Kenics mixer shows the highest pressure loss in comparison to the other studied mixers which is also displayed in Fig. 8. The pressure drop of the Kenics mixer is on average 60 % higher than the pressure loss of the non-obstructed pipe flow for all measured flow rates.

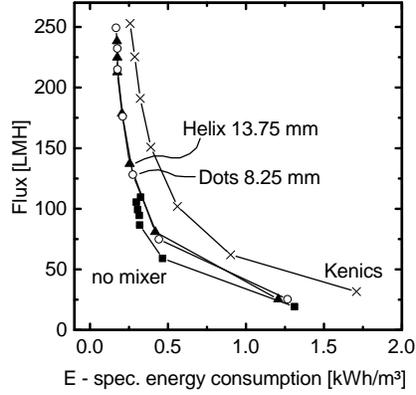

Figure 9: Flux versus specific energy consumption for selected systems with static mixer.

Furthermore, it can be concluded from Fig. 8 that the differences in pressure drop are more pronounced for higher flow rates. The lowest analyzed flow rate of 40 $L/h$ corresponds to a Re number of 2353 for the empty tube, hence, the pressure loss was measured in the transient region for the empty tube. For the systems with mixer, the flow might already be turbulent due to the flow obstruction by the mixers.

Yet, the insertion of static mixers does not only increase the pressure drop and hence lead to a higher energy requirement, but static mixers significantly increase the permeate flux. For this reason, the specific energy consumption is calculated according to Eq. 1:

$$E = \frac{\dot{Q} \cdot \Delta p}{J \cdot A} + TMP = \frac{\dot{Q} \cdot (p_F - p_R)}{J \cdot A} + TMP \qquad (1)$$

Here, $\dot{Q}$ denotes the volumetric feed flow rate and $\Delta p$ corresponds to the pressure drop along the membrane. The latter is calculated by averaging the difference of feed and retentate pressure throughout the complete fouling experiment. In Eq. 1, the first term describes the energy required to pump the fluid along the membrane whereas the second term corresponds to the energy needed to filtrate. Both terms are related to the obtained amount of permeate. Thus, the specific energy consumption describes the total pump energy required per $m^3$ of permeate.

Fig. 9 shows the permeate flux as a function of specific energy consumption for selected filtration systems. It is clearly visible that the application of static mixers increases the obtained permeate flux when the same specific energy is invested. Hence, the membrane area required to obtain a certain amount of permeate can be reduced by applying static mixers without increasing the necessary specific energy. As can be seen from Fig. 9, the insertion of a Kenics static mixer leads to the highest increases in permeate flux for the investigated systems. All mixer geometries which are based on the twisted tape show similar flux vs. specific energy profiles which are in between the Kenics mixer and the membrane without mixer. For comparison, only the results for the helical mixer with pitch length 13.75 $mm$ and for the twisted tape with "dotted" structure and a pitch length of 8.25 $mm$ are shown.

Furthermore, Fig. 9 illustrates that the advantages of static mixers are especially found in the region of low specific energy consumption which corresponds to high permeate fluxes (cf. Eq. 1). These high permeate fluxes are - due to severe fouling - not reachable in the system without static mixer or at least not reachable at economically justifiable expenses.

## 4. Conclusion

Various geometries of turbulence promoters - or static mixers - have been applied to mitigate fouling in a tubular ceramic membrane. Using 3D-printing techniques to fabricate the mixers gave more freedom in designing their geometry. Most mixer geometries were based on a twisted tape structure. The diameter of two mixers was varied over their length; either the diameter decreased continuously and increased sharply or the diameter was varied in a sinusoidal shape. In addition, a Kenics static mixer was applied as turbulence promoter in the flow channel of the membrane.

All investigated static mixers showed a significant increase in flux compared to a membrane filtration using no static mixer. The permeate flux was at least 53 % higher in the systems with mixer when compared to the system without mixer. The most significant flux improvement was achieved by the Kenics static mixer with 140 %, followed by the twisted tape geometries with constant diameter. Applying static mixers with varying diameter increased the permeate flux less than the mixers with constant diameter. The smaller effect of the mixers with changing diameter might be attributed to short-cut streams evolving due to the smaller mixer diameter.

In addition to the increase in permeate flux, the specific energy consumption was considered which relates the energy required to pump and filtrate the fluid to the produced amount of permeate. The application of static mixers increases the required pumping energy as the mixers obstruct the flow and increase the pressure drop along the membrane. However, they also enlarge the permeate flux. As the increase in flux is significantly higher than the increase in pressure drop, all static mixers decrease the specific energy consumption compared to the system without mixer. Thus, it can be concluded that introducing static mixers in the flow channel of tubular membranes is an efficient measure to mitigate fouling.

**Acknowledgement**

The authors acknowledge financial support from ERA-NET SUSFOOD-CEREAL project, No. 031A431B, by the German Federal Ministry of Education and Research. This project has also received funding from the European Research Council (ERC) under the European Union's Horizon 2020 research and innovation program (grant agreement no. 694946).